\documentclass[sigconf, nonacm]{acmart}
\usepackage[colorinlistoftodos]{todonotes}
\usepackage{subcaption}

\newcommand\vldbdoi{XX.XX/XXX.XX}
\newcommand\vldbpages{XXX-XXX}
\newcommand\vldbvolume{19}
\newcommand\vldbissue{1}
\newcommand\vldbyear{2026}
\newcommand\vldbauthors{\authors}
\newcommand\vldbtitle{\shorttitle}

\newcommand\vldbpagestyle{plain}

\begin{document}
\title{Accelerating Presto with GPUs}

\author{Daniel Bauer}
\author{Luis Garc\'es-Erice}
\author{Deepak Majeti$^\ast$}
\author{Zolt\'an Arnold Nagy}
\author{Sean Rooney}

\affiliation{%
  \institution{IBM Research Europe,$^\ast$ IBM Data \& AI}
}
\author{Greg Kimball}
\author{Devavret Makkar}
\author{Todd Mostak}
\author{Karthikeyan Natarajan}
\affiliation{%
  \institution{NVIDIA}
}

\begin{abstract}

We describe how we extended Presto to be GPU-aware. We focus on two critical challenges: efficiently moving data from storage to GPU operators, and enabling data exchange between operators without leaving GPU memory
even when a query is distributed.
To guide our design, we conducted a series of initial experiments in which we executed queries derived from the TPC-H benchmark on a multi-GPU cluster using NVIDIA's C++ cuDF data-frame library, and measured how different architectures and configurations influenced performance. 
We show how these insights inform our extensions to Presto, detailing the architectural changes required to integrate GPU execution into the existing Presto framework. Our initial evaluation demonstrates substantial cost/performance (up to 6$\times$) improvements over CPU Presto on standard analytical benchmarks. Our code is available as part of open-source Presto/Velox, and we have started to use it to run customer production workloads.

\end{abstract}

\maketitle

\pagestyle{\vldbpagestyle}
\begingroup\small\noindent\raggedright\textbf{PVLDB Reference Format:}\\
\vldbauthors. \vldbtitle. PVLDB, \vldbvolume(\vldbissue): \vldbpages, \vldbyear.\\
\href{https://doi.org/\vldbdoi}{doi:\vldbdoi}
\endgroup
\begingroup
\renewcommand\thefootnote{}\footnote{\noindent
This work is licensed under the Creative Commons BY-NC-ND 4.0 International License. Visit \url{https://creativecommons.org/licenses/by-nc-nd/4.0/} to view a copy of this license. For any use beyond those covered by this license, obtain permission by emailing \href{mailto:info@vldb.org}{info@vldb.org}. Copyright is held by the owner/author(s). Publication rights licensed to the VLDB Endowment. \\
\raggedright Proceedings of the VLDB Endowment, Vol. \vldbvolume, No. \vldbissue\ %
ISSN 2150-8097. \\
\href{https://doi.org/\vldbdoi}{doi:\vldbdoi} \\
}\addtocounter{footnote}{-1}\endgroup

\section{Introduction}

GPU acceleration of analytics has been researched for over a decade~\cite{10.1145/3318464.3380595,sharma2024comprehensiveoverviewgpuaccelerated,10.1007/978-3-319-01863-8_25,yogatama2025rethinkinganalyticalprocessinggpu}, yet commercial SQL engines remain largely CPU-bound. We close that gap by extending Presto with GPU support, letting existing deployments adopt GPU acceleration without migrating to a new system.

GPU performance gains over CPU are primarily due to the extremely high memory bandwidth of modern GPUs, often more than an order of magnitude greater than that of comparable CPUs~\cite{10.1145/3318464.3380595,yogatama2025rethinkinganalyticalprocessinggpu}.
This, coupled with the fact that analytics data processing workloads are generally memory~\cite{10.1145/3318464.3380595}
rather than compute-bound, results in GPUs outperforming CPUs on these workloads \textit{if} the GPUs can be properly exploited.
The authors of~\cite{aramburú2025theseusdistributedscalablegpuaccelerated} state that most prior work on GPUs concentrates on raw compute speed yet overlook the cost of moving data to and from device memory. We focus on this aspect of building a performant data processing system for GPUs. 

Presto~\cite{10.1145/3589769} is a complete SQL-enabled distributed data processing system, with a single coordinator supporting multiple worker nodes. The coordinator plans queries and the workers execute the query plan in parallel across the cluster.
Many commercial offerings have been built using Presto, for example, AWS Athena~\cite{aws-athena-presto}.
Although standard Presto workers are written in Java, \textit{Prestissimo} (also known as Presto Native) is a compatible implementation of a Presto worker that uses Velox C++ libraries~\cite{10.14778/3554821.3554829} as the core execution engine. 
For the rest of this paper we will use Presto to mean a Java Presto coordinator
running multiple Presto Native workers.

NVIDIA's cuDF~\cite{rapids:18} library manipulates data frames using
GPUs. It uses the Apache Arrow~\cite{10.1145/3527199.3527264} columnar format to load representations of data into memory
and allows standard data operations, e.g.~filtering, joining, aggregation, etc.
The library offers high-level primitives that exploit a single GPU's SIMT (Single Instruction, Multiple Threads) model to allow the parallel execution of the same operation
e.g.~filtering, across all or parts of a cuDF table loaded into GPU memory.
GPU libraries also exist for moving data: KvikIO~\cite{kvikio:25} for handling data transfers from storage to GPUs and UCX~\cite{openucx-website} for transfers between GPUs.
In this paper we show how we integrated support for GPUs into Presto
using these existing GPU libraries as our starting point. Our implementation is fully integrated into the open-source Presto and Velox repositories, i.e.~it is not simply a research prototype, and is available for third party use. Because we have extended open-source Velox, any Velox-based engine, e.g.~Spark/Gluten~\cite{gluten2024}, inherits the same GPU support. This demonstrates how GPU acceleration can be retrofitted into existing analytical ecosystems without requiring migration to new systems.

First we describe our initial "bare-bones" experimental set-up, detailing the 
software and hardware components of our cluster that allowed us to experiment with the key drivers of query performance. During these experiments, we did not use Presto
or any other high-level data processing system, allowing us to completely control the query plans, schedules, data format, exchange protocols, compression algorithms etc. 
We demonstrate a large gap in performance between
what our GPU infrastructure permits and what existing data processing systems
actually achieve. 
This work led us to conclude that any performant system must
bring data efficiently into the GPU's memory and retain it there
during the lifetime of the query.

Then we report how we have extended the Velox architecture to exploit GPUs, creating GPU-aware versions of standard Velox operators using the cuDF libraries and show how these can be 
transparently made available to existing workloads. 
These operators read parquet efficiently into GPU memory and minimize synchronization during the execution of a pipeline of operators.

We explain how we retained the data working set in GPU memory for distributed queries
by extending Presto's inter-worker exchange protocol to be GPU-aware
and report that when data can be contained in the GPU memory 
of the Presto cluster that we record a 20$\times$ increase in performance over the standard Presto exchange protocol for certain exchange heavy queries.
Our overall query cost was up to 6$\times$ cheaper than what we achieved with CPU-only Presto for the same performance as measured on standard benchmarks.

\section{Understanding Performance Limits}
\label{sec:set-up}

\begin{figure*}[htbp]
    \centering
     \includegraphics[width=0.8\linewidth]{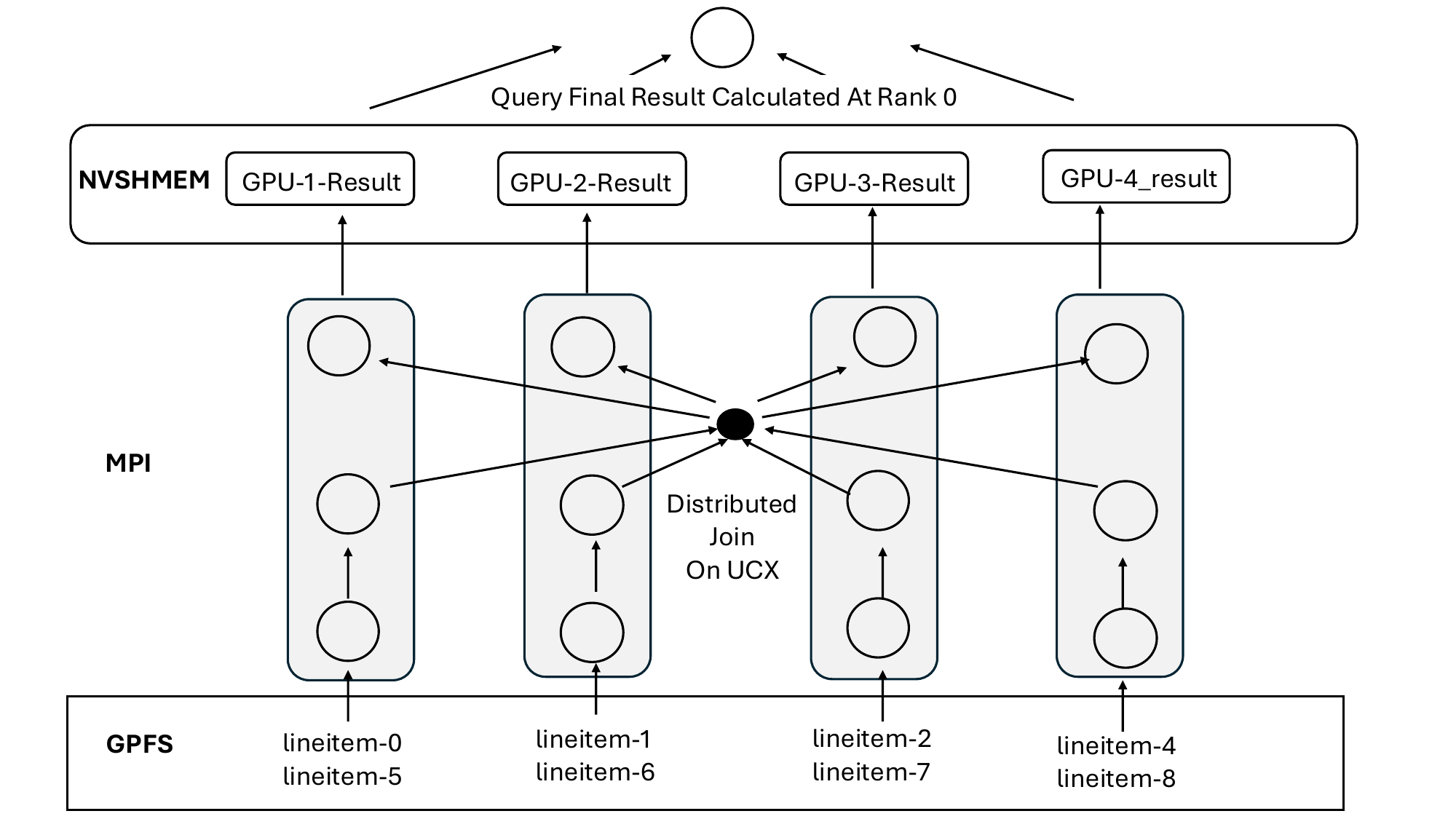}
        \caption{Overview of Software Architecture for Motivating Experiments}
        \label{fig:overview}
\end{figure*}

Before extending Presto, we conducted foundational experiments to validate three hypotheses about GPU query performance:
\begin{enumerate}
    \item \textbf{Minimize CPU-GPU data movement}: Reading data directly into GPU memory, avoiding CPU staging, is critical for performance.
    \item \textbf{Retain data in GPU memory}: Data should remain in GPU memory throughout query execution, including across operator boundaries.
    \item \textbf{GPU-native inter-node communication}: For distributed queries, data exchange must occur directly between GPUs without CPU memory staging.
\end{enumerate}

These experiments did not use Presto or any other high-level query engine---they represent a separate, minimal system built solely to establish performance bounds and guide our design. The actual Presto integration is described in Section~\ref{sec:presto-integration}. Here, our goal was to determine what the hardware could achieve with minimal software complexity, establishing a performance target for our Presto implementation.

Our evaluation was performed on a representative set of queries derived from the TPC-H benchmark~\footnote{We use "TPC-H-like" because official TPC-H results require formal auditing and compliance certification. Our queries are functionally identical to TPC-H, but we have not undergone the official audit process, so the results should not be directly compared to certified TPC-H benchmarks.} that we implemented directly on the CUDA environment using
cuDF. Our goal was to determine the maximum obtainable performance for those queries on 
10 TB of data (SF=10,000) by handcrafting the execution of the needed operators and the query plan. We excluded spilling from GPU-to-CPU memory when the query failed due to GPU memory exhaustion but searched for solutions that ensured that
the working data set did not exceed the total
amount of available memory i.e.~we wanted to determine a lower bound on the latency that could be achieved 
with the infrastructure available to us when no data had to be spilled from GPU memory. We note that current generation GPUs, such as NVIDIA B300 Ultra or AMD MI350X, have 288 GB of memory~\cite{yogatama2025rethinkinganalyticalprocessinggpu} meaning
that it is feasible to run medium size analytics workloads entirely from memory.

\subsection{Experimental Infrastructure}
\label{subsec:infrastructure}

We ran our experiments on two Supermicro AS-4124GO-NART systems, each with 256 CPU cores, 2TB RAM, and 8$\times$A100 80GB GPUs (16TB/sec aggregate memory bandwidth per server), connected to a 160TB GPFS storage cluster with 170 GB/s read throughput. We used KvikIO~\cite{kvikio:25} with GPU Direct Storage (GDS) for direct storage-to-GPU data transfers. For process coordination, we created an MPI process per GPU~\cite{MPI-1994}, with each worker processing a fraction of the input data. Processes exchange partial results via NVSHMEM~\cite{NVSHMEM:25}, which provides a global symmetric memory space accessible by all processes, and perform distributed joins using UCX~\cite{openucx-website} following the approach in~\cite{Gao2021ScalingJT}. Figure~\ref{fig:overview} illustrates this architecture. This minimal setup---without a query planner, coordinator, or high-level exchange protocol---allowed us to establish performance bounds with minimal software overhead.

\subsection{Data Format Bottlenecks}
\label{subsec:data-formats}

We initially generated TPC-H data in Parquet format, but found that reading it with the cuDF 25.04 version's Parquet reader was 10$\times$ slower than the theoretical I/O rate possible on the infrastructure described in Section~\ref{subsec:infrastructure}.

Parquet's hierarchical structure---with nested metadata at file, row-group, and page levels~\cite{parquet_1_0}---requires continuous interpretation during reads. This metadata complexity, combined with interleaved data and decode operations, prevents GPUs from fully exploiting memory bandwidth; others have made the same observation~\cite{helfman2024nimble}.

To ensure that the complexity of interpreting metadata did
not slow down the overall system, we created a simple file format 
similar to Apache Arrow's persistent representation, modifying the TPC-H tool dbgen~\cite{electrum_tpch_dbgen} to generate compact data-sets. 
We ignored nulls as there are none within TPC-H
and divided columns into distinct files, allowing us to use the file name to encode the
necessary minimal metadata, i.e.~the name of the column, its type
and the compression algorithm used.
We encoded strings in two columns, the first containing the
actual string data and the second the string offsets. 
We also experimented with partition strategies for the data with different chunk sizes
for the tables at different scale factors.
We decided not to allow the reading of only parts of a file as this would have required additional metadata, e.g.~about the the location of data boundaries. As we wished to experiment with the influence of
the size of the data unit processed on query performance we generated different chunk size for the same scale factor, i.e.~breaking a specific column into 64, 128, 256 etc. parts.
So a specific table is made up a set of files whose cardinality is the number of columns $\times$ the number of chunks.
We measured the I/O read rate when reading
TPC-H tables generated using our format into GPU memory as cuDF tables
at 95\% of the maximum theoretical throughput.
At SF-1000, the \textit{lineitem} table is roughly 400 GB in our own format, meaning we are
reading at 22 GB/s.

To be clear: we are not proposing this format as an alternative to Parquet. The purpose of this experiment is to \textit{quantify the performance gap} between what the hardware can achieve (95\% of theoretical I/O throughput) and what current formats deliver (10$\times$ slower with Parquet). The order of magnitude gap establishes a concrete performance target and demonstrates that data format overhead is a first-order concern for GPU query processing, not merely an implementation detail.

While more recent cuDF versions (e.g., 25.10) read Parquet more efficiently, the complex metadata structure makes it inherently difficult to fully exploit GPU memory bandwidth.
Designing GPU-friendly data formats is an avenue for future research; however, exabytes of existing data are already stored in formats like Parquet. Our Presto integration therefore uses Parquet (see Section~\ref{subsec:implementing-gpu-aware-ops}), accepting current format limitations while benefiting from GPU acceleration elsewhere in the query pipeline.

\subsection{Query Performance Results}

\begin{table}[h!]
\centering
\small
\begin{tabular}{|l|r|r||l|r|r|}
\hline
\textbf{Q} & \textbf{Parts} & \textbf{Time(s)} & \textbf{Q} & \textbf{Parts} & \textbf{Time(s)} \\
\hline
Q1 & 256 & 8.2 & Q13 & 512 & 12.8 \\
Q2 & 64 & 4.2 & Q14 & 128 & 10.5 \\
Q6 & 128 & 5.7 & Q16 & 64 & 4.2 \\
Q9 & 128 & 77.0 & Q17 & 128 & 18.7 \\
Q10 & 128 & 25.9 & Q20 & 64 & 19.6 \\
Q11 & 64 & 4.3 & & & \\
\hline
\end{tabular}
\caption{TPC-H-like query performance at SF=10,000 (10TB). Parts = number of data partitions.}
\label{tab:tpch-perf}
\end{table}

Table~\ref{tab:tpch-perf} shows the \textit{lowest} execution time observed for a subset
of TPC-H-like queries at SF=10,000 (i.e. 10TB) from multiple runs across a set of
different configurations. The 12 queries were chosen in order to include a wide set of
different operations. For example, \textit{Q9} requires high performance joins
while \textit{Q1} requires efficient aggregation~\cite{boncz:2014}.
As a comparison \textit{Q1} using Presto Native on the same hardware
i.e.~with 256 CPU Cores and 2TB RAM and the same optimized storage
 but without using A100s, is 15$\times$ slower, even when it uses caching,
data skipping and query optimization. In Section~\ref{subsubsec:scale-up}, we show that our Presto/Velox integration achieves comparable performance to these handcrafted baselines on newer GPU hardware.

Since each query required manual implementation in this bare-bones setup, we selected a representative subset covering the key operator types: simple aggregation (Q1, Q6), multi-way joins (Q2, Q9, Q10), and combinations of filters, joins, grouping, and ordering (Q11, Q13, Q14, Q16, Q17, Q20). Our goal was not to achieve a "best score" on TPC-H, but to establish a performance baseline across diverse operator patterns.
Each run is cold as there is no data cache in the system, i.e.~data is read from storage using GDS. There is no metadata store where statistics about the distribution of data across tables can be stored
to optimize query plans and there is no capacity to skip reading data by a consideration of file metadata.

Larger chunks always
gave better results for all scale factors, but at some chunk size the GPU ran out of memory and a smaller chunk needed to be used, so the value shown is always
for the smallest number of chunks for which the query completed.
In Table~\ref{tab:tpch-perf} one can see that this varies across the queries ranging from 64 to 512 partitions.
An intelligent query planner would attempt to determine dynamically the largest chunk size 
that would fit into total GPU memory. 

We note that approaches that worked well on smaller data sets failed as we increased the scale factor. 
It was also the case that approaches that worked well at a high scale factor were often slower than other approaches at lower scale factors. 
This suggests that the query planner should choose the operator implementation
based on both the expected input and the available resources. Possibly
operators should be assembled from sub-operators in a way similar to that described in ~\cite{modularis2021} using different patterns according to resources constraints. For example, the distributed inner join defined in~\cite{Gao2021ScalingJT} failed as soon as the working set exceeds the total available memory on the GPUs.
At scale factor SF=10,000 the lineitem table is 3.5~TB in our proprietary format meaning that the table by itself cannot be loaded into the totality of our cluster's GPU memory. In consequence, a solution that works at lower scale factors will fail at higher ones.
  For larger scale factors we used \textit{late materialization}~\cite{Stonebraker2005Cstore} to constrain GPU memory usage at the cost of additional table reads:
  \begin{enumerate}
      \item project each partition to join-keys only, reducing memory footprint;
      \item execute the distributed join on key-only tables;
      \item reread tables from storage to add missing columns via local join.
  \end{enumerate}

  Since rows may materialize on any GPU after the distributed join, step (3) requires joining against the entire table, not just the locally assigned partition.

As all the workers must join against the entire table a further pattern is having
individual workers read a partition of the table in (2) and then broadcast via NVSHMEM
the part of the table read to all other workers. This reduces contention at the storage layer.

The data query processing GPU platform
Theseus~\cite{aramburú2025theseusdistributedscalablegpuaccelerated}, developed by Voltron
Data, Inc.\ (which has since ceased operations), is a system built from the ground up for
GPU acceleration.  It may be considered the state-of-the-art in terms of published
performance figures.  The Theseus paper reports a total of 230 seconds for a 16 node GPU
cluster for SF=10,000. The paper does not break down the performance per query, but
roughly speaking our minimal set-up achieves a similar order of magnitude.  Theseus was a
general purpose system offered as a commercial product, while our experiment is completely
handcrafted. As such, they are not comparable in terms of practical usage. However, the
performance we report is achieved simply by not getting in the way of what the hardware
allows. The components in the experimental set-up use generic frameworks such as cuDF and
UCX, so in principle a standard distributed query engine could achieve the same results if
it can move data efficiently into GPU memory and \textit{keep it there} during the entire
execution of the query across a multi-node GPU cluster.

These experiments validated our hypotheses: the performance we report is achieved by moving data efficiently into GPU memory and keeping it there during the entire execution of the query. We now describe how we extended Presto following these principles, designing our additions to be as transparent as possible so that existing Presto installations can benefit from GPU acceleration with minimal changes.

\section{GPU Integration in Presto Native}
\label{sec:presto-integration}

Guided by the three hypotheses validated in Section~\ref{sec:set-up}, we extend Presto/Velox along two axes: Section~\ref{subsec:implementing-gpu-aware-ops} swaps standard Velox operators for cuDF-backed equivalents so data stays on the GPU within a pipeline, and Section~\ref{subsec:exchanging-data} adds \textit{UcxExchange} for GPU-to-GPU data exchange across workers. Sections~\ref{subsection:results}--\ref{subsec:scaling} evaluate the combined system.

\subsection{Overview of Velox}
\label{subsection:overview}

Velox is an open-source, high-performance, vectorized execution engine designed to unify data processing across various compute engines. 
A Velox pipeline is a chain of execution units called operators that process data in a vectorized, pipelined, and multi-threaded fashion. Examples of operators are relational operations such as \textit{TableScan}, \textit{Filter}, \textit{Join}, etc.
In order to scale, all operations performed on a pipeline are asynchronous.
A driver thread asks a specific operator if it can perform work, and if not, it
returns immediately, using \textit{futures} to preserve the state of the operator. 
The drivers of a pipeline are work conserving, as they will never idle in a blocked
state if there is work to do.
In the pipeline each operator processes vectors of data and passes them to the next
operator. Despite the name, a \textit{RowVector} is not a single row but a batch of rows (typically thousands), enabling efficient vectorized processing. Execution is pipelined: each batch is processed and passed downstream as it becomes available, rather than waiting for all data to accumulate. This streaming-of-batches model balances GPU efficiency (which benefits from larger batches) with pipeline concurrency (stages operate concurrently on different batches).

External to Velox a Presto query planner generates a plan from a SQL query
and divides the plan into distinct Presto stages, decomposing each stage into a plan fragment.
Velox workers on distinct nodes receive a plan fragment, each creating
a task to manage it. The task creates multiple pipelines from the plan fragment and assign driver threads to execute them. A pipeline is finished when all available input data has been advanced to the end of the pipeline. A pipeline starts with a data source, 
either from storage (\textit{TableScan} operator) or from another Velox worker (via an \textit{Exchange} operator). A pipeline finishes by either writing data or sending it to other pipelines. When these pipelines are remote
this is done via a \textit{PartitionedOutput} operator that uses an exchange protocol to transfer data to multiple remote
\textit{Exchange} operators.

Velox does not define how the exchange protocol is implemented.
This is the responsibility of the distributed processing system in which it is used,
e.g. Presto or Spark.
Presto's existing exchange is based on HTTP. In \textit{HttpExchange} on the upstream side, the results of
a task are serialized into pages, where a page is the smallest unit of transmission. The size
of a page is configurable, typically in the order of a few megabytes. The downstream client uses
a request/reply protocol to fetch the serialized pages. Pages are only requested when the
receiving process has sufficient memory for receiving. 

Integrating GPU execution into Velox posed several non-trivial challenges beyond simply replacing CPU operators with GPU equivalents:
\begin{enumerate}
    \item \textbf{Asynchronous execution model}: Velox's asynchronous, futures-based execution model must be preserved when offloading work to GPUs. We designed \textit{CudfVector} to encapsulate both data and a CUDA stream, enabling multiple pipelines to issue overlapping GPU commands without blocking CPU driver threads.

    \item \textbf{Cross-pipeline synchronization}: When data flows between pipelines (e.g., during joins), streams must be synchronized to ensure data consistency while maximizing concurrency. Our design synchronizes source streams only at pipeline boundaries.

    \item \textbf{Expression compilation}: Velox's expression framework compiles expressions at plan creation time. We implemented a translation layer that converts Velox \textit{TypedExpr} trees to \textit{CudfExpression} trees, with a hybrid approach that prefers cuDF's fused AST execution for performance while falling back to standalone functions for unsupported operations.

    \item \textbf{Streaming operators without native support}: cuDF lacks true streaming versions for some operators (e.g., \textit{groupby}). We implemented workarounds such as concatenation-based streaming aggregation that respects memory limits while maintaining correctness.

    \item \textbf{GPU-native exchange protocol}: Presto's HTTP-based exchange requires CPU memory staging. We designed \textit{UcxExchange} from scratch using UCX, implementing tag-based rendezvous for GPU-to-GPU transfers, active messaging for handshakes, and flow control to prevent memory exhaustion---all while integrating with Velox's asynchronous execution model.
\end{enumerate}

\begin{figure*}[htbp]
    \centering
    \begin{minipage}{0.45\textwidth}
         \raggedleft
        \includegraphics[width=0.95\linewidth]{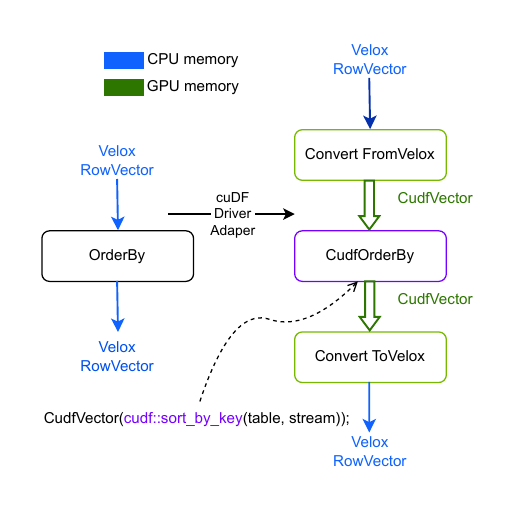}
        \caption{Velox Translation of CPU Operators to GPU Operators and data memory location}
        \label{fig:translator}
    \end{minipage}\hfill
    \begin{minipage}{0.45\textwidth}
        \raggedright
        \includegraphics[width=0.95\linewidth]{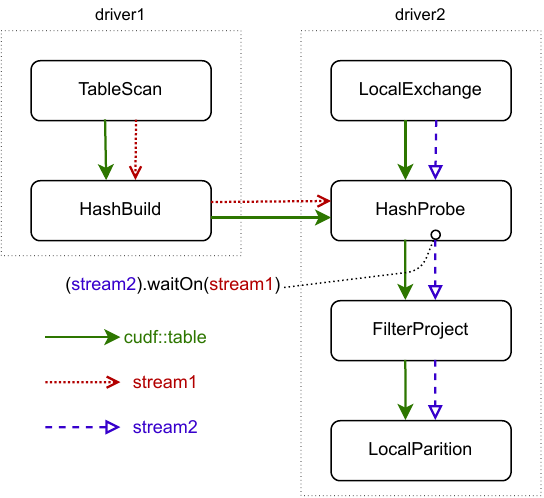}
        \caption{Velox CudfVector Data Flow -- table with stream}
        \label{fig:dataflow_cudfvector}
    \end{minipage}
\end{figure*}

We extended Velox to support the exploitation of GPUs using software libraries built on cuDF.
In our implementation each GPU is managed by an independent Velox worker, i.e.~a given Velox GPU-aware operator uses exactly one GPU.
Having a one-to-one mapping between Velox workers and GPUs
greatly simplifies the task of synchronization and memory management.

Velox supports a feature called driver adaption that allows the transformation of a pipeline of operators before execution. 
We used this feature such that during the construction
of the pipeline from a plan fragment, we replace standard Velox operators
with their cuDF equivalent when these are available.

When all operators in a pipeline have a cuDF version, then
the data is entirely processed in the GPU.
For example a GPU-aware \textit{TableScan} reads from Parquet files into GPU memory
creating a \textit{CudfVector} that can be used by subsequent operators, e.g.~a \textit{CudfFilterProject} 
that does the actual filtering on the GPU without the CPU being involved,

In order to allow backward compatibility with existing Velox
workload, we allow operators that are GPU-aware to be interleaved with those that aren't. Specific
conversion operators \textit{CudfToVelox} and \textit{CudfFromVelox} are implemented which do the conversions, copying to/from GPU/CPU memory. When the translators recognizes that
a non GPU-aware operator is required the appropriate conversion
operators are added. 
Both Velox and cuDF are Arrow-compatible, so often the interoperation is a simple copy with no transformation or encode/decode required.
Note that this a pragmatic choice so that the system is immediately usable even if not every Velox operator has a GPU-aware equivalent. 

Figure~\ref{fig:translator} illustrates an example of a translation showing
the original pipeline containing an \textit{OrderBy} operator and how it is modified to add a GPU aware
version of that operator \textit{CudfOrderBy} as well as the means to convert to/from \textit{CudfVectors} and Velox \textit{RowVectors}.

\subsection{Implementing GPU-aware Velox operators}
\label{subsec:implementing-gpu-aware-ops}

At the time-of-writing approximately 50\% of Velox operators have GPU-aware implementations, covering the most frequently used operators. This coverage is sufficient to execute all 22 TPC-H queries and all 99 TPC-DS queries entirely on GPUs without falling back to CPU execution. Extending coverage to the remaining operators is straightforward engineering work with no fundamental technical barriers---cuDF provides GPU implementations for all standard SQL operations.

Avoiding CPU fallback is critical: when a non-GPU-aware operator is encountered, data must be copied from GPU to CPU memory via \textit{CudfToVelox}, processed on the CPU, and then copied back via \textit{CudfFromVelox}. As demonstrated throughout this paper, this memory transfer overhead can dominate query execution time, negating the benefits of GPU acceleration. This is why achieving complete operator coverage is essential for production deployments.

Our GPU-aware operators use a \textit{CudfVector} data structure that wraps a \textit{cudf::table} (cuDF's Arrow-compatible columnar format in GPU memory) along with a CUDA stream for asynchronous execution. Each Velox pipeline uses a single stream, with synchronization occurring only at pipeline boundaries when data is exchanged. This design allows GPU operators to return immediately to Velox's CPU drivers without blocking, preserving Velox's asynchronous execution model. Figure~\ref{fig:dataflow_cudfvector} illustrates how multiple pipelines can execute concurrently using different streams.

For data ingestion, we extended Velox's \textit{HiveConnector} with a GPU-aware \textit{CudfHiveConnector} that uses cuDF's GPU-accelerated Parquet reader to load data directly into GPU memory, taking advantage of GDS when available.

For expression evaluation, we translate Velox's \textit{TypedExpr} trees into equivalent \textit{CudfExpression} trees. cuDF offers two execution modes: standalone functions (one kernel per operation) and AST evaluation via \textit{cudf::compute\_column} (fused operations, fewer memory writes). We implement a mixed translation system that prefers AST for performance but falls back to standalone functions for unsupported operations.

For aggregations, Velox's \textit{HashAggregation} operator supports multiple modes (\textit{Partial}, \textit{Intermediate}, \textit{Final}, \textit{Single}) to enable distributed execution with exchanges between stages. Since cuDF lacks a true streaming groupby with dynamic hash table expansion, our \textit{CudfHashAggregation} uses a concatenation-based approach: each batch is partially aggregated, concatenated with previous results, and re-aggregated until a size threshold triggers output emission.


\subsection{Exchanging Data Between GPUs in a Cluster}
\label{subsec:exchanging-data}

Prior to our work the only exchange mechanism available in Presto was 
the HttpExchange. While it is possible to
continue using this with the operators described in Section~\ref{subsec:implementing-gpu-aware-ops} doing so 
requires a conversion between \textit{RowVector} and \textit{CudfVector} before
and after every transmission between Presto workers.
The moving of data between CPU and GPU has a dramatic negative effect on overall performance.

We implemented new operators
that allow the direct exchange of cuDF tables between GPUs across a cluster,
independently of whether they are on the same physical machine or are connected via the network.
This approach allows data to be exchanged between GPU-aware Velox workers without requiring the transfer to the host memory. 
When the worker is GPU-aware, the Velox operators that use exchange
between workers \textit{PartitionedOutput} and \textit{Exchange} are replaced with GPU-aware operators 
\textit{CudfPartitionedOutput} and \textit{UcxExchange}. These
interact with our new \textit{UcxExchange} protocol.

\subsubsection{UCX}
Our exchange protocol builds on UCX, the communication framework introduced in Section~\ref{sec:set-up}. UCX automatically selects optimal transport mechanisms---NVLink for same-machine GPU transfers, RDMA for cross-machine, or TCP as fallback---based on endpoint capabilities and topology. We use UCX's \textit{tagSend}/\textit{tagRecv} for asynchronous data transfer and active messaging for the initial handshake to establish rendezvous tags.

\subsubsection{UcxExchange Protocol}

In production workloads with queries much more complicated than TPC-H, e.g.~having hundreds of stages, spanning 40+ tables, and containing dozens of joins, we observed that highly skewed joins sometimes resulted in the creation of many millions of small vectors. This in turn saturated the cuDF libraries, resulting in massive performance degradation. In consequence, we added optional vector compaction as part of the exchange: small vectors are merged before transmission to maintain efficient batch sizes for GPU processing. This demonstrates both the need for real-world workloads for testing and some of the particularities of the GPU that a scheduler might need to account for.

The Presto coordinator assigns each task a partition to consume from remote tasks, identified by a URL (remote split). It is not known ahead of time how much data is associated with a split. The consumer reads the stream of \textit{CudfVectors} from the remote source one at a time until notified that there is no more data. The size of the individual \textit{CudfVectors} being processed is governed in part by the size of the chunks of data being read in the initial \textit{TableScan}; in our configuration we typically set this to be as large as possible while not causing out-of-memory errors, e.g.~4~GB.

\textit{CudfVectors} are transferred as two parts: CPU-resident metadata (schema and size) and GPU-resident packed data (cudf columns gathered into contiguous memory). The receiver initiates a handshake via active messaging to establish a unique rendezvous tag, then issues asynchronous \textit{tagRecv} calls---first for metadata to determine allocation size, then for the packed GPU data. An empty metadata message signals transfer completion.

\begin{figure*}[htbp]
    \centering
     \includegraphics[width=1.0\linewidth]{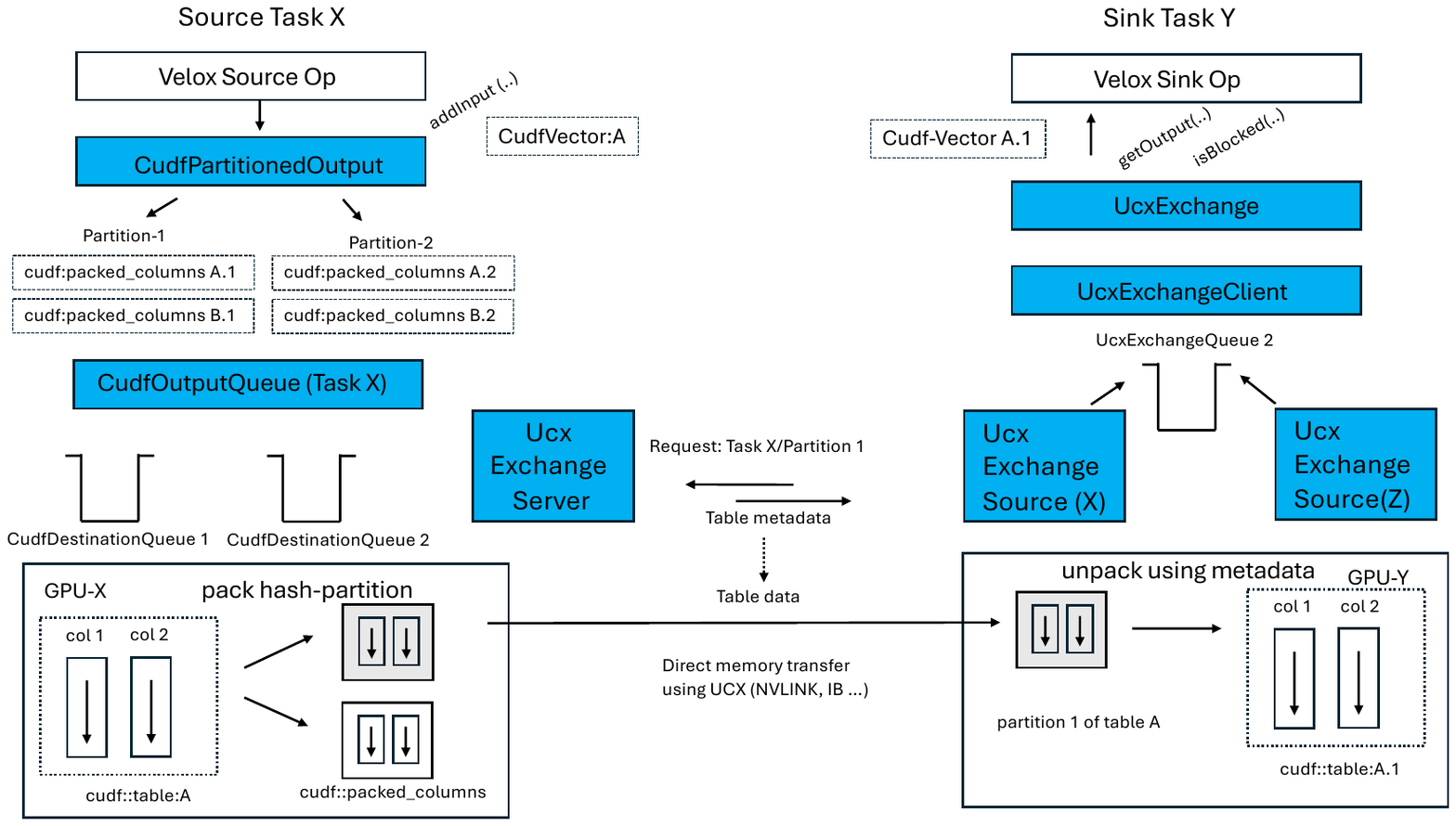}
        \caption{UcxExchange Architecture}
        \label{fig:exchange-architecture}    
\end{figure*}

Figure~\ref{fig:exchange-architecture} illustrates the exchange architecture. Source task \textit{X} uses \textit{CudfPartitionedOutput} to partition data into queues; sink task \textit{Y} creates \textit{UcxExchangeSource} instances to pull data from remote workers via UCX. All operations are asynchronous, with tasks potentially requesting data before sources are created.

Since Presto stages execute concurrently and data is temporarily duplicated during transfers, the working set must fit in cluster GPU memory. We implement basic flow control, blocking sends when queues exceed thresholds. A complete solution requires global resource-aware scheduling as proposed by~\cite{aramburú2025theseusdistributedscalablegpuaccelerated, 10.14778/3749646.3749710}.
However making the scheduler aware of the resources in the system 
and creating the plan accordingly would involve changing the Presto query planner and was beyond the scope of our current work. We discuss this further in Section~\ref{sec:future-work}.

\subsubsection{Presto Integration}

We now describe
how \textit{UcxExchange} is used within a Presto worker.
A Presto worker may simultaneously be a data source and a data target 
within the execution of the same SQL query offering data from its tasks to downstream task and at the same time requesting data from other upstream tasks. The worker contains both a server component and one or more client components. 
All UCX operations are applied within a \textit{UCXContext} object that retains the
communication context e.g.~gathered transport capabilities, the memory manager etc. A \textit{UCXWorker} is the actual engine that establishes communications with other \textit{UCXWorkers}.
There can only be one \textit{UCXContext} per process and only a single \textit{UCXWorker} node per context.

Multi-threading on top of the \textit{UCXWorker} is supported but poses some challenges with respect to message ordering, as it is up to the receiving threads to decide
for which thread a specific message is intended.
\textit{UcxExchange} implements a single-threaded UCX-based communication layer within one worker that drives both the sending and receiving of data. The single thread
that drives the communication is not a bottleneck as the actual
transfer of the data is done by UCX independently 
of this thread. The communicator is best thought of as a simple control mechanism that coordinates UCX.

The communicator contains the main loop that drives the communication between senders and receivers. It initializes UCX and encapsulates the UCX context, worker and listener. The communicator creates the UCX listener for incoming connection requests and also registers an active message handler for handling the initial request for data from receivers. This handshake contains sufficient information to uniquely identify the receiver.

At each worker the communicator maintains a list of remote UCX endpoints. These endpoints are shared by senders and receivers such that between any two workers only one UCX connection per-direction exists. These connections are established on the first exchange between the two workers. It is costly to establish a UCX connection as topology information is gathered and exchanged by both sides, so although reusing connections between different
exchanges is more complex it significantly reduces the latency.

Communication elements --- either senders or receivers --- register with the communicator and put themselves into a work queue that gets serviced in a upcall in the communicator’s thread. For each sender and receiver in the work queue, the communicator calls the corresponding \textit{process} method once.
The communicator calls \textit{progress} on the UCX worker to advance the UCX communication. This indirectly also invokes the upcalls in senders and receivers.

The principal challenge with creating the \textit{UcxExchange} was combining the programming model of UCX into that of Velox.
Other than the hooks to initialize the communicator
nothing else was modified within the Presto workers/coordinator. The \textit{UcxExchange} can be activated via a Presto configuration option when starting a
worker. Certain other options may be set which determine which GPU the worker is using and various configuration options for UCX, but otherwise the usage of the GPU during the exchange is completely transparent to the user of the system.



\subsection{Results: Query Performance}
\label{subsection:results}

With \textit{UcxExchange} enabled, the total time for all 22 TPC-H queries at SF=1000 is 93 seconds, compared to 828 seconds using the default \textit{HttpExchange}---a more than 8$\times$ speedup. The geometric mean per query improves from 22.2s to 3.8s (6$\times$ speedup). This difference is entirely due to avoiding the CPU/GPU data movement bottleneck.

\begin{table}[h!]
\centering
\small
\begin{tabular}{|l|l|}
\hline
\textbf{Component} & \textbf{Configuration} \\
\hline
Server & Supermicro AS-4124GO-NART \\
CPUs & 256 cores, 2TB RAM \\
GPUs & 8 $\times$ NVIDIA A100 (80GB each) \\
GPU Interconnect & NVLink (intra-node) \\
Storage & GPFS, 160TB, 170 GB/s read \\
\hline
Presto & Presto Native (Prestissimo) \\
Velox & With cuDF GPU operators \\
cuDF & Version 25.04 \\
UCX & For GPU-to-GPU transfers \\
\hline
Benchmark & TPC-H-like (22 queries) \\
Scale Factors & SF=100 to SF=10,000 \\
Workers & 1 to 8 (one per GPU) \\
Runs per config & 10 (cold start, no caching) \\
\hline
\end{tabular}
\caption{Experimental Setup Summary}
\label{tab:experimental-setup}
\end{table}

\begin{figure*}[t]
    \centering
        \includegraphics[width=1.0\linewidth]{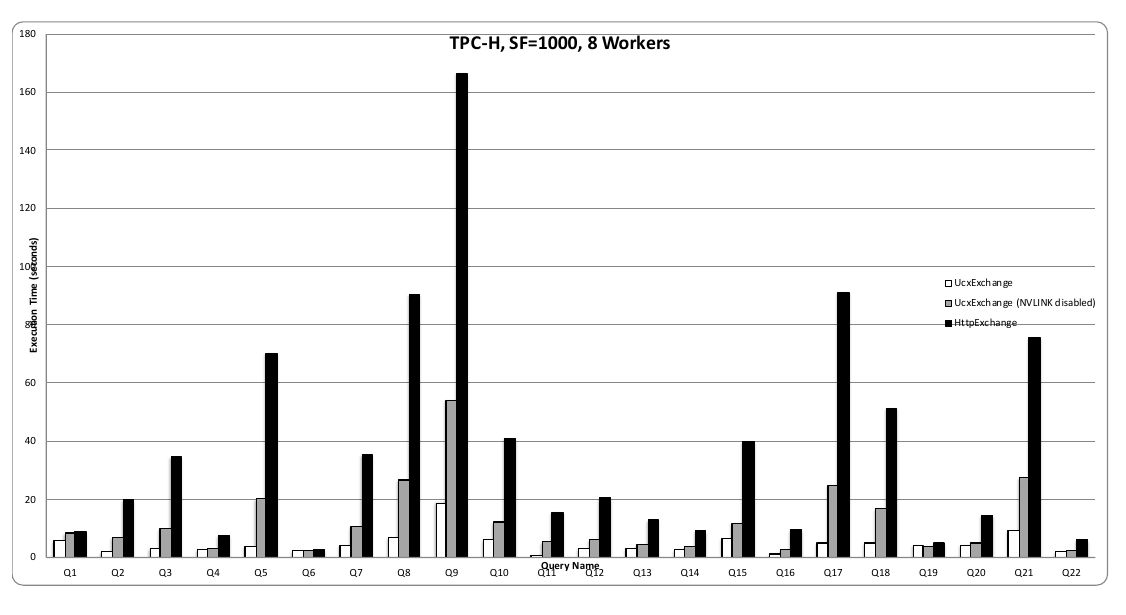}
        \caption{Query execution time (seconds) for all 22 TPC-H queries at SF=1000 on 8$\times$A100 GPUs, comparing HttpExchange (CPU-staged) vs UcxExchange (GPU-direct) vs UcxExchange without NVLink.}
        \label{fig:exchange-noexchange-perf}
\end{figure*}

Table~\ref{tab:experimental-setup} summarizes our experimental setup. The A100 GPU used was released in 2020. Using this three-generation-old GPU demonstrates that our approach does not depend on cutting-edge hardware to achieve significant performance improvements, and that users with existing cheaper GPU infrastructure can benefit immediately. For reference we also report performance on systems based on the Blackwell B200 in Figure~\ref{fig:sf10000-scaleup}.
We use TPC-H-like queries at different scale factors to measure the performance
of Presto GPU. As all the Velox operators required by Presto for executing TPC-H have a GPU-aware equivalent we can run all queries without
leaving the GPU. We run these tests on one of the 8$\times$A100 GPU machines described in~Section~\ref{sec:set-up}. We run 8 Presto workers on the machine, each worker assigned to one of the GPUs.
The Velox \textit{AsyncDataCache} is disabled, meaning every run is a cold run, reading the required data from storage for \textit{each} query.

To show the cost of moving data to/from the CPU
we give results with and without our \textit{UcxExchange},
i.e.~forcing the data transfer to CPU memory on every cross-worker communication.
In addition, we run the \textit{UcxExchange} with NVLink disabled to show what portion of the speed up
is due to it. Note that when NVLink is disabled on our infrastructure,
UCX chooses RDMA as the best protocol for exchange between GPUs, meaning data is passed between GPUs via the NICs.
The largest scale factor at which all Presto queries run with 8$\times$A100 GPUs (640 GB RAM)
is SF=1000, so we report on results for that configuration.

Figure~\ref{fig:exchange-noexchange-perf} compares query performance for all 22 queries. The results shown are averages over 10 runs each from a cold start. The coefficient of variation is below 10\% on all the reported measurements.

Queries that require large volumes of data to be exchanged due to
large joins, e.g.~\textit{Q9}, run more than 20$\times$ faster with \textit{UcxExchange} enabled. As we would expect, queries which exchange minimal amounts of data do not show any benefit and sometimes incur a small penalty.
For example, \textit{Q1} involves only transmitting 4 rows; as such, the use
of the UCX is overkill. Potentially an intelligent scheduler could determine
what parts of a plan could benefit from GPUs and schedule accordingly,
i.e.~have plans that account for heterogeneous workers with/without GPUs. This would require extensions to the exchange to allow efficient GPU/CPU transmission.
We discuss this idea further in Section~\ref{sec:future-work}.

Disabling NVLink reduces \textit{UcxExchange} performance to 269 seconds, but is still 3$\times$ faster than the standard \textit{HttpExchange}, showcasing the benefits of avoiding HTTP transmission and directly transferring data between GPUs.

\begin{figure}[htbp]
    \centering
     \includegraphics[width=1.0\linewidth]{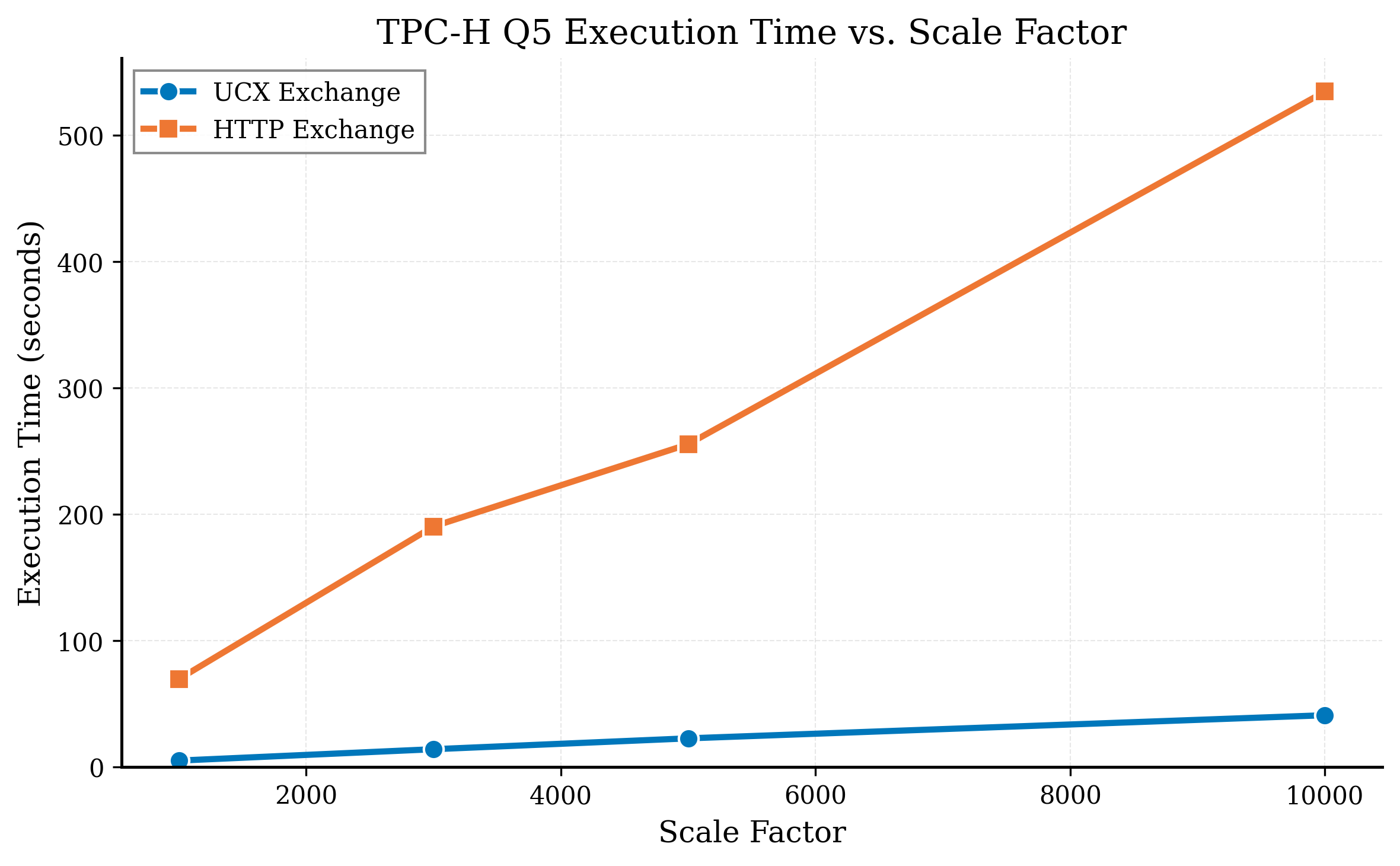}
        \caption{Q5 execution time (seconds) on 8$\times$A100 GPUs across scale factors SF=1000--10000, comparing HttpExchange vs UcxExchange. UcxExchange maintains $>$10$\times$ speedup at all scale factors.}
        \label{fig:exchange-noexchange-sf-perf}
\end{figure}

In order to show how performance varies across scale factor, we chose a query that was join-heavy, but ran on our infrastructure
up to scale factor SF=10,000.
Figure~\ref{fig:exchange-noexchange-sf-perf} shows how the execution of \textit{Q5} on 8 workers varies with increasing scale factors SF=(1000...10,000) with both \textit{UcxExchange} and \textit{HttpExchange}. It can be seen that the ratio between
their performance remains stable with the scale factor and is never below 10$\times$.

\begin{figure*}[htbp]
    \centering
     \begin{subfigure}[b]{0.48\linewidth}
        \centering
        \includegraphics[width=\linewidth]{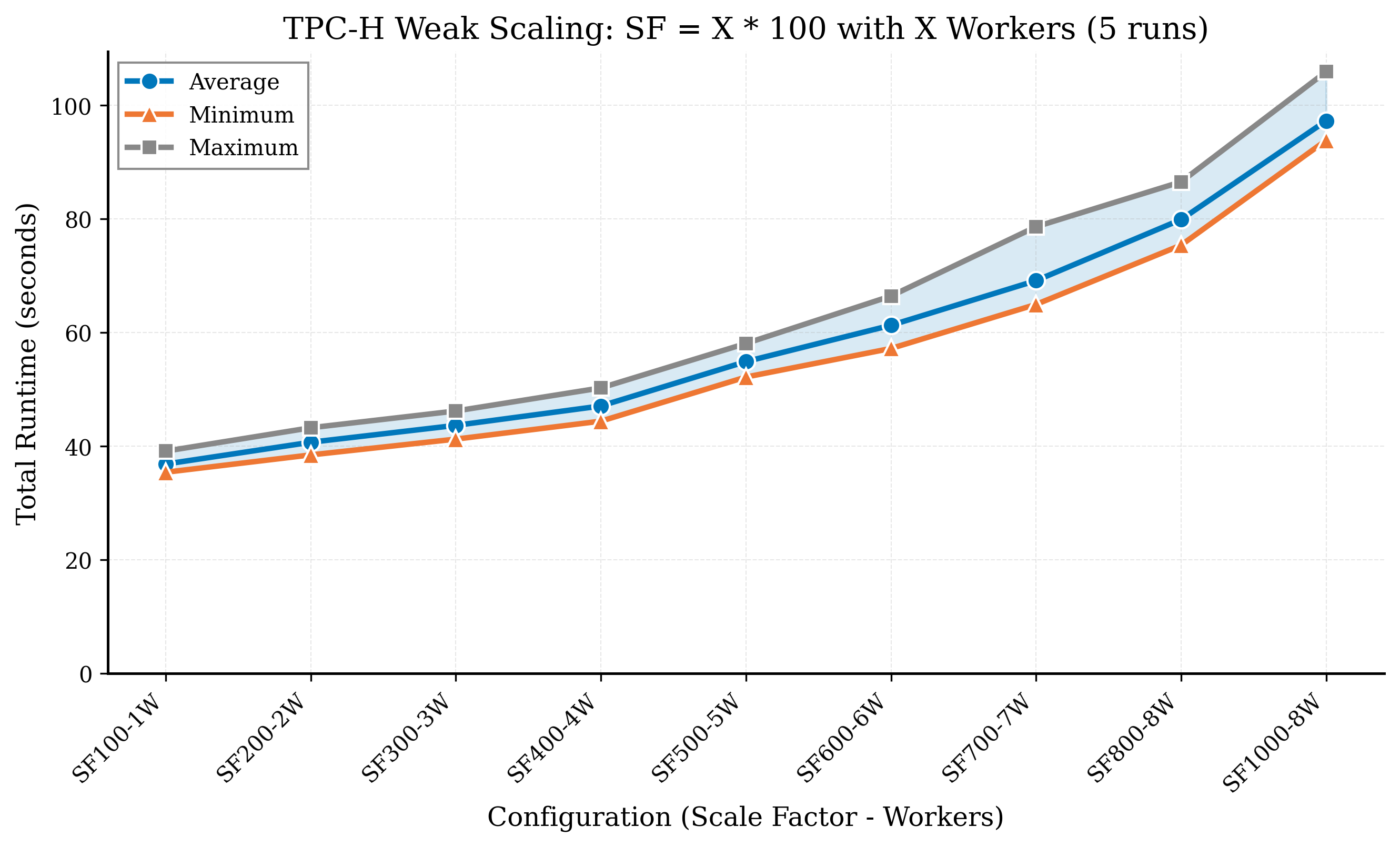}
        \caption{Total benchmark time (seconds) for all 22 TPC-H queries at SF=100--1000 with 1--8 A100 GPUs (one worker per GPU). Shows min, max, and average across 5 runs.}
        \label{fig:tpch-sfx00-wx-5-runs}
    \end{subfigure}
    \hfill
    \begin{subfigure}[b]{0.48\linewidth}
        \centering
        \includegraphics[width=\linewidth]{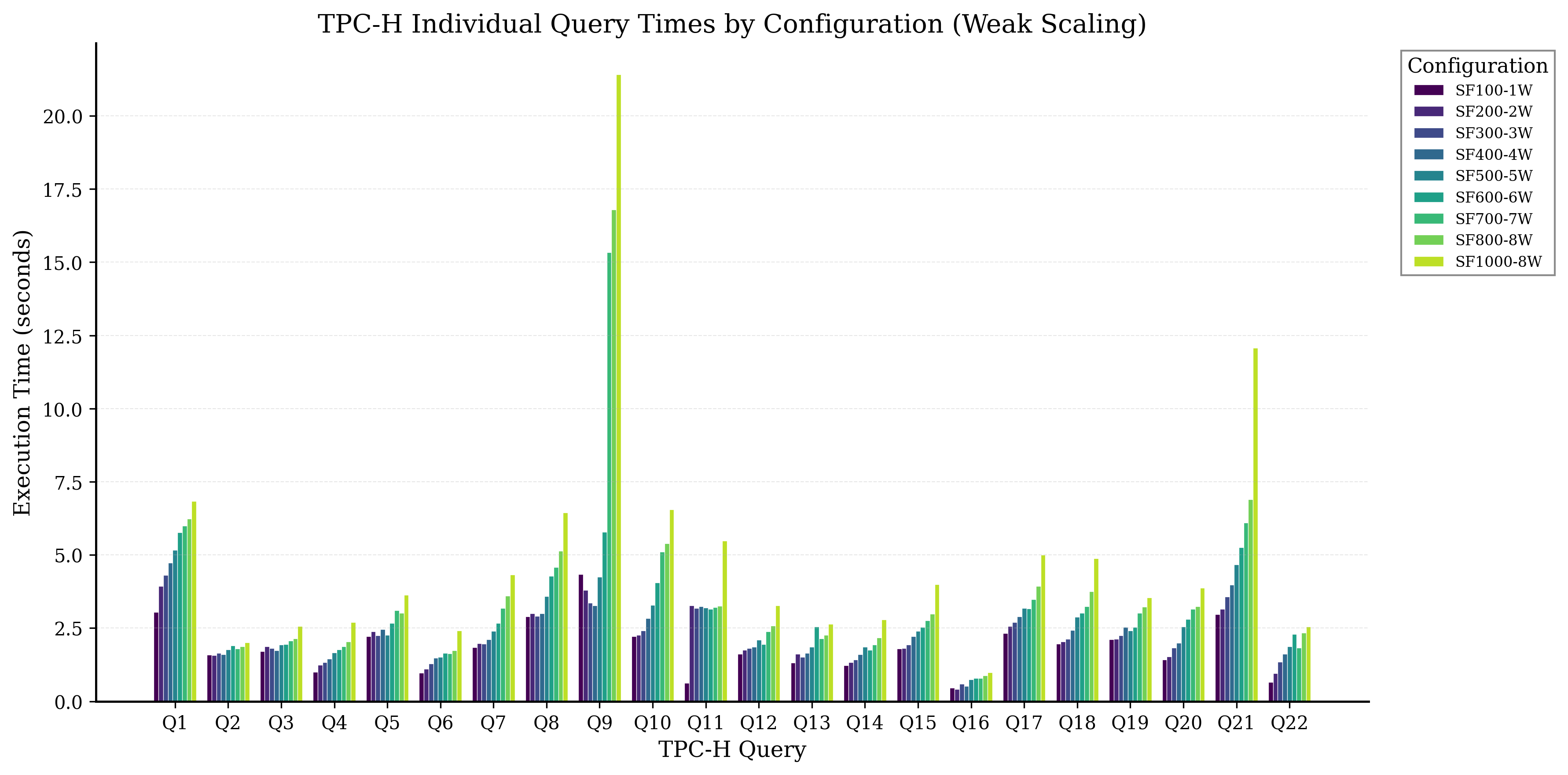}
        \caption{Per-query execution time (seconds) across SF=100--1000 with 1--8 A100 GPUs. Q9 and Q21 (join-heavy) show the largest increases with scale.}
        \label{fig:time_for_sf_queries_tpch}
    \end{subfigure}
        \caption{Weak-scaling analysis: TPC-H benchmark on A100 GPUs where data size and worker count increase proportionally (SF=100/1 GPU to SF=1000/8 GPUs).}
        \label{fig:weak-scaling}    
\end{figure*}

Figure~\ref{fig:weak-scaling} measures Presto's ability to maintain a constant execution time as the problem size and number of processors are both increased proportionally, i.e.~the performance of running the \textit{entire} benchmark at SF=100 on 1 worker compared with SF=200 on 2 workers etc.
In Figure~\ref{fig:tpch-sfx00-wx-5-runs} it can be seen that the time to complete the TPC-H-like benchmark increases as X=(scale factor \& number of workers) increases, but the increase is sub-linear: processing 8 times as much data with 8 times as many workers doubles the processing time. Further increasing the data size to SF=1000 with 8 workers shows a more pronounced effect on the time to complete the benchmark.

Figure~\ref{fig:time_for_sf_queries_tpch} shows how each query's runtime changes at different scalings/workers. \textit{Q9} contributes the most to the deviation from perfect scalability, with longer runtimes for larger scale factors/workers. \textit{Q21} shows similar behavior. Both queries are very join heavy.
Our expectation is that we can get closer to linear scaling by better exploiting driver parallelism during joins within Velox.

\subsection{Scaling to SF=30,000}
\label{subsec:scaling}

The results in Section~\ref{subsection:results} are limited to SF=1000 due to the 640~GB total GPU memory of our 8$\times$A100 configuration. We demonstrate two approaches to scale beyond this limit: using GPUs with more memory (scaling up), and adding additional servers (scaling out).

\subsubsection{Scaling Up}
\label{subsubsec:scale-up}

\begin{figure*}[htbp]
    \centering
    \includegraphics[width=1.0\linewidth]{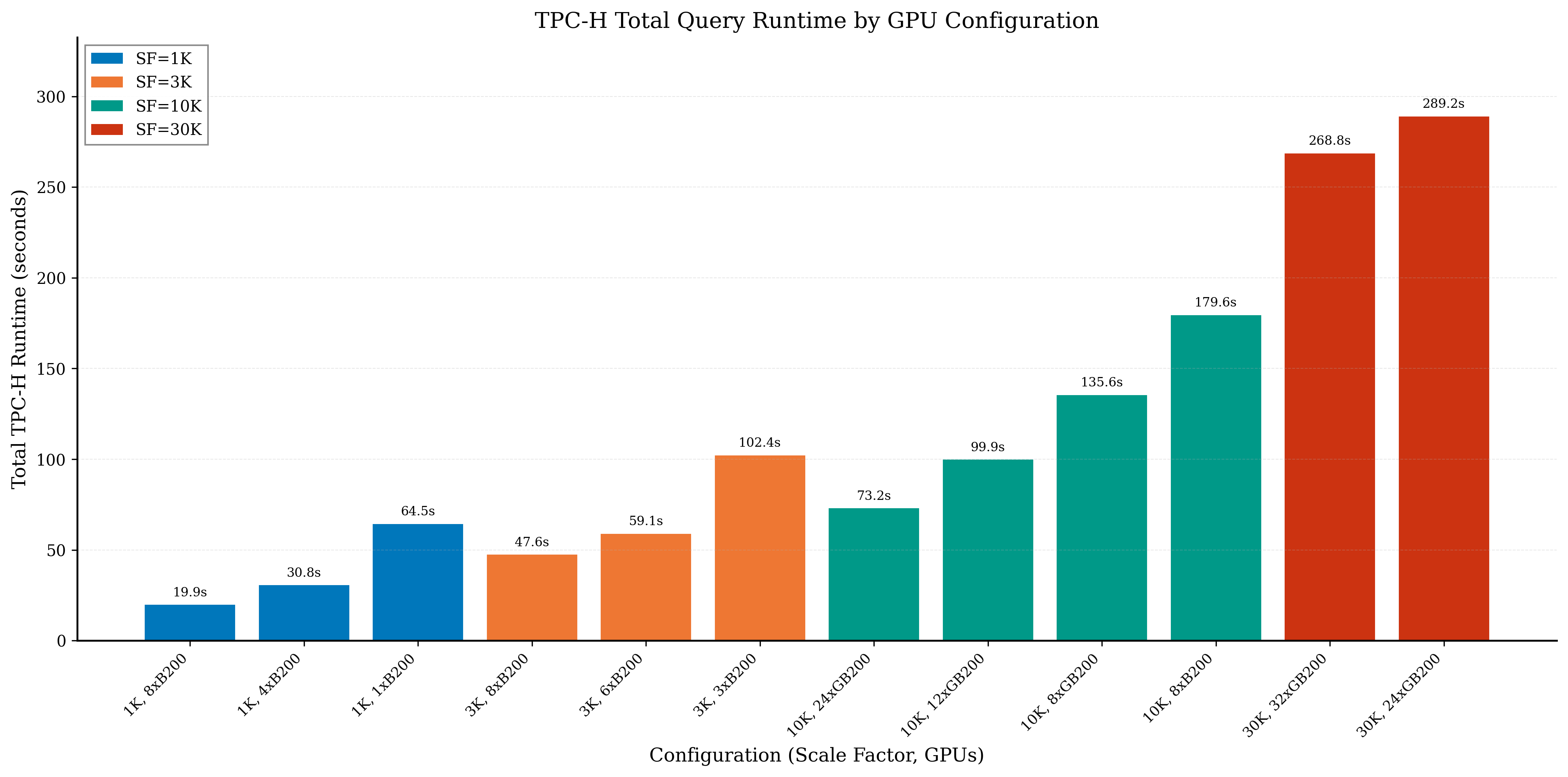}
    \caption{Total TPC-H runtime by GPU configuration on single servers with modern GPUs. Scale factors range from 1K to 30K (30TB), with GPU counts from 1 to 40.}
    \label{fig:sf10000-scaleup}
\end{figure*}

The results in Section~\ref{subsection:results} used 8$\times$A100 GPUs (80~GB each, 640~GB total), limiting us to SF=1000. To demonstrate what is achievable with current-generation hardware, we evaluated performance on single servers with more modern GPUs: NVIDIA DGX systems with B200 GPUs (192~GB each) and NVL72 systems with GB200 GPUs. Figure~\ref{fig:sf10000-scaleup} shows the total TPC-H runtime across different scale factors and GPU configurations.

Figure~\ref{fig:sf10000-scaleup} demonstrates the benefit of newer GPU hardware. At SF=1K with 8 B200 GPUs, total runtime is 19.9s compared to 93s with 8 A100 GPUs (Section~\ref{subsection:results})---a 4.7$\times$ improvement from faster GPUs alone. Strong scaling is evident: at SF=1K, increasing from 1 to 8 B200 GPUs reduces runtime from 64.5s to 19.9s (3.2$\times$ speedup).

Comparing to our barebones experiments (Table~\ref{tab:tpch-perf}), Q1 at SF=10K took 8.2s on 16$\times$A100 GPUs (1.28~TB total memory). With Presto/Velox, Q1 achieves 6.3s on 8$\times$B200 (1.54~TB), 4.7s on 12$\times$GB200 (2.3~TB), and 2.6s on 24$\times$GB200 (4.6~TB). These newer GPUs have significantly more memory and compute capacity; the results show that our Presto integration scales effectively with improved hardware.

The larger memory of modern GPUs enables much larger scale factors on a single server. At SF=10K (10TB), 8 B200 GPUs complete the benchmark in 179.6s, while 24 GB200 GPUs on an NVL72 achieve 73.2s. The system scales to SF=30K (30TB)---30$\times$ larger than what was possible with A100s---completing the full TPC-H benchmark in approximately 4.5 minutes with 24--40 GB200 GPUs. These results demonstrate that our Presto GPU integration benefits directly from GPU hardware improvements transparently to users.

\subsubsection{Scaling Out}
\label{subsubsec:scale-out}

 \begin{table}[htbp]
  \centering
  \caption{TPC-H SF3K Performance (10 Workers, AWS g7e)}
  \begin{tabular}{|l|r|}
  \hline
  \textbf{Run Type} & \textbf{Time (s)} \\
  \hline
  Hot   & 108 \\
  Cold  & 190 \\
  \hline
  \textbf{Ratio (Cold/Hot)} & \textbf{1.77x} \\
  \hline
  \end{tabular}
  \label{tab:tpch-sf3000-aggregate}
  \end{table}

To demonstrate horizontal scaling we run a Presto cluster on AWS increasing the number of single GPU VMs (g7e.8xlarge) until we can run the SF=3000. Each of the RTX PRO GPUs have 96 GB RAM. We can run the benchmark with 10 Presto workers. The servers have each 256 GB of RAM.
Table~\ref{tab:tpch-sf3000-aggregate} shows the cold (from S3) and hot runs from cache. The 1.77$\times$ difference between cold and hot runs highlights the importance of I/O performance; the cache is large enough to hold the working set for all queries. Comparing to our SF=1000 results on 8$\times$A100 GPUs (93s, 640~GB total GPU memory, Section~\ref{subsection:results}), we process 3$\times$ the data in only 16\% more time (108s hot) by scaling to 10 GPUs (960~GB total GPU memory).

  \textbf{Summary of key findings:} Our evaluation demonstrates three main results: (1) \textit{UcxExchange} provides up to 20$\times$ speedup over \textit{HttpExchange} for exchange-heavy queries by keeping data in GPU memory; (2) even without
  NVLink, direct GPU-to-GPU transfers via RDMA achieve 3$\times$ improvement over CPU-mediated exchange; and (3) the system exhibits reasonable scaling behavior—processing 8$\times$ more data on 8$\times$ more workers results in only 2$\times$
  longer execution time. The system scales both vertically (using GPUs with more memory) and horizontally (adding more servers).

\subsection{Comparison GPU v CPU}

Direct performance comparison between GPU and CPU systems is inherently problematic because the hardware architectures differ fundamentally. The appropriate metric for comparison is cost-effectiveness: the dollar cost to achieve a given level of performance. This approach has been adopted by other GPU analytics systems including Theseus~\cite{aramburú2025theseusdistributedscalablegpuaccelerated} and Sirius~\cite{yogatama2025rethinkinganalyticalprocessinggpu}. We therefore compare cost/performance using AWS as our cost model~\cite{AWS_Pricing}.

\begin{figure*}[htbp]
    \centering
    \includegraphics[width=0.9\linewidth]{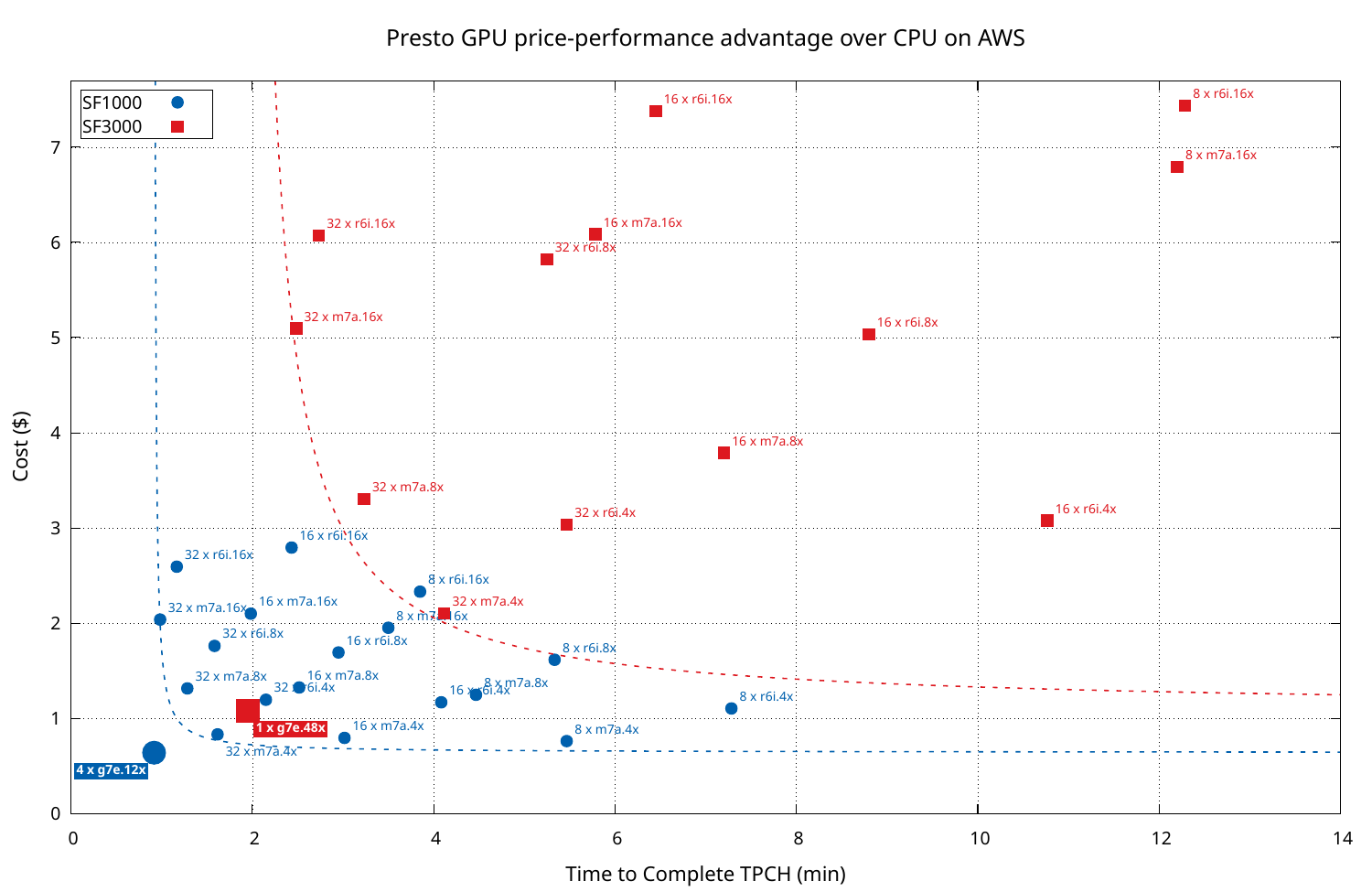}
    \caption{Price-performance Presto GPU advantage on AWS.}
    \label{fig:price-performance}
\end{figure*}

Figure~\ref{fig:price-performance} compares cost and runtime for TPC-H on AWS (warm runs with caching from S3) across GPU and CPU configurations at SF=1K and SF=3K. GPU configurations are highlighted: 4$\times$g7e.12xlarge (8 GPUs total) for SF=1K and 1$\times$g7e.48xlarge (8 GPUs) for SF=3K, both using NVIDIA RTX 6000 PRO GPUs without NVLink. CPU configurations span r6i and m7a instance families at various sizes.

The dashed Pareto frontiers reveal diminishing returns for CPU clusters: reducing runtime requires disproportionately higher cost. In any tested configuration, CPUs do not match GPU cost/performance. We quantify price-performance as the cost$\times$time product (lower is better). At SF=1K, the GPU achieves 0.59 (\$0.65$\times$0.9~min) versus 1.35 for the best CPU configuration---a 2.3$\times$ advantage. At SF=3K, the GPU achieves 1.87 (\$0.85$\times$2.2~min) versus 11.84 for the best CPU---a 6.3$\times$ advantage. This advantage grows with scale factor as GPU memory bandwidth benefits larger datasets.


This results are consistent with other recent studies: Theseus reports up to 4$\times$ better performance than Databricks Photon at cost parity~\cite{aramburú2025theseusdistributedscalablegpuaccelerated}, and Sirius demonstrates 7--8$\times$ better cost efficiency on TPC-H and ClickBench~\cite{yogatama2025rethinkinganalyticalprocessinggpu}.
Note that the cost difference is likely to increase as previous generations of GPUs are replaced with newer ones for model training and as a result are less in demand~\cite{yogatama2025rethinkinganalyticalprocessinggpu}.

\subsection{Limitations}

Our evaluation has several limitations that should be considered when interpreting the results:

\begin{itemize}

    \item \textbf{Synthetic workload}: We use TPC-H-like queries which, while standard for benchmarking, may not fully represent production workloads. Real workloads exhibit different data distributions, query patterns, and concurrency characteristics. We have run real customer workloads on GPU-aware Presto and have seen significant performance benefits for non-trivial queries involving more than 50 joins and 254 Presto stages executing on 2.4 TBs of data.

    \item \textbf{Memory constraints}: Our current implementation requires the working set to fit in GPU memory. Queries that exceed available GPU memory fail rather than spilling to CPU memory, limiting applicability to very large datasets without cluster scale-out. We are actively developing
    memory managers that allow the efficient spilling from GPU to CPU memory.

    \item \textbf{No query planner customization}: The Presto query planner is not yet GPU-aware. An intelligent planner could further improve performance by considering GPU memory constraints and communication costs when generating query plans. One advantage of using an existing ecosystem
      is that advances in query optimization, e.g.~\cite{HBO2024}, can immediately be exploited. However, whether there is a need for customization
      of planners to make them GPU-aware is work in progress.
\end{itemize}

Despite these limitations, our results demonstrate substantial performance improvements for queries that fit within the system's constraints, and the architecture provides a foundation for addressing these limitations in future work.

\section{Related Work}
\label{sec:related-work}

GPU-accelerated query processing has been an active research area for over a decade. We survey the most relevant systems and position our work within this landscape.

\textbf{GPU Database Systems.}
HeavyDB (formerly MapD/OmniSciDB)~\cite{HeavyDB2017} is one of the earliest commercial GPU-accelerated databases, designed from the ground up for GPU execution with its own storage format and query engine.
BlazingSQL~\cite{BlazingSQL2020} is an open-source GPU SQL engine built on cuDF and Apache Arrow---the same libraries we use---but operates as a standalone system rather than extending an existing distributed query engine.
PG-Strom~\cite{PGStrom2016} takes a different approach by extending PostgreSQL with GPU acceleration for scans and joins, demonstrating the value of GPU integration into existing systems.

\textbf{Distributed GPU Query Processing.}
Theseus~\cite{aramburú2025theseusdistributedscalablegpuaccelerated} is a recent distributed GPU query platform optimized for data movement, reporting strong TPC-H results at large scale factors. Unlike our work, Theseus is built from scratch rather than extending an existing system.
Sirius~\cite{yogatama2025rethinkinganalyticalprocessinggpu} explores the economics of GPU analytics, arguing that GPU cost-effectiveness is improving rapidly---a trend that motivates our work on GPU-accelerating Presto.
The work on scaling joins to thousands of GPUs~\cite{Gao2021ScalingJT} addresses distributed join algorithms using NVSHMEM, which we leverage in our preliminary experiments.

\textbf{Hybrid CPU-GPU Execution.}
Several systems explore hybrid execution strategies. Li et al.~\cite{10.14778/3749646.3749710} present techniques for scaling GPU databases beyond GPU memory by performing filtering on CPU before GPU processing.
HyPE~\cite{Heimel2013HyPE} pioneered hardware-oblivious query processing that can dynamically choose between CPU and GPU execution.
Yuan et al.~\cite{Yuan2013YDB} studied the trade-offs of GPU query processing, identifying when GPU execution is beneficial.
Breß et al.~\cite{10.1007/978-3-319-01863-8_25} explored the design space of GPU-aware database architectures, identifying key trade-offs we also encountered.

Table~\ref{tab:related-work} compares our approach with these systems. Our work differs in three key aspects: (1) we extend an existing production system (Presto/Velox) rather than building from scratch, enabling existing Presto users to benefit from GPU acceleration; (2) we implement a GPU-native exchange protocol (UcxExchange) that avoids CPU memory staging during distributed execution; and (3) our implementation is contributed to open-source, making it available for production use.

\begin{table}[h!]
\centering
\small
\begin{tabular}{|l|c|c|c|}
\hline
\textbf{System} & \textbf{Extends} & \textbf{GPU} & \textbf{Open} \\
                & \textbf{Existing} & \textbf{Exchange} & \textbf{Source} \\
\hline
HeavyDB & No & Yes & Partial \\
BlazingSQL & No & Yes & Yes \\
PG-Strom & Yes (PG) & No & Yes \\
Theseus & No & Yes & No \\
\textbf{This work} & \textbf{Yes (Presto)} & \textbf{Yes} & \textbf{Yes} \\
\hline
\end{tabular}
\caption{Comparison with GPU query processing systems.}
\label{tab:related-work}
\end{table}

\section{Future Work}
\label{sec:future-work}

In Section~\ref{sec:set-up} we motivated the need for a GPU-aware query
planner to generate plans that take full advantage of
the hardware: for example, optimizing the strategy to
keep the working data set in GPU memory, distinguishing between queries
and/or sub-queries that should be executed on the GPU and those that
can be executed on the CPU, using locality between workers to
optimize data transfer, etc.
Velox CPU execution possesses a mature memory management infrastructure that facilitates resource sharing across tasks and manages out-of-memory events by triggering data spilling. As we address memory management for Velox-cuDF, we have identified two complementary approaches operating at different levels: (1) automatic spilling with CUDA managed memory, where cuDF's unified memory managers allocate memory across CPU/GPUs, allowing queries that exceed GPU memory; and (2) explicit spilling within the application, enabling trading memory between concurrent queries. These approaches are complementary: unified memory handles within-query overflow while application-level spilling manages cross-query resource sharing. However, unified memory managers must work correctly with all libraries in the stack, including UCX, to avoid performance degradation from unnecessary CPU memory indirection.

An ideal GPU-aware query planner/coordinator
would use the distinct capabilities and resources
of the many components, e.g.~KvikIO, UCX, etc described in this paper, to make better decisions about the placement of queries.
In preparation for this we have already extended Presto workers to gather and report topology
information to the coordinator. Even the NUMA topology of a single server can
influence the ideal placement of a query plan.
We are also investigating hybrid CPU/GPU execution strategies that can dynamically route operators based on data characteristics and resource availability. Whether ultimately making analytical data processing GPU-aware would be better achieved
by extending an existing system or building an entirely
new one remains an open question~\cite{yogatama2025rethinkinganalyticalprocessinggpu}.

In Section~\ref{subsec:data-formats} we motivated why more \textit{GPU-friendly} data formats would be advantageous. Another alternative that we are actively investigating is to create a level of indirection between the reading of data
from storage and its transmission to the GPUs. The data reading can be
parallelized such that a large number of cheap CPUs are used to load
the deserialized data into memory and only then make
it available to the expensive GPU.
We are extending our Exchange system to efficiently
transmit data from many CPUs simultaneously into the GPU. This is especially advantageous when the storage system is slow but highly scalable, such as Amazon S3.

\section{Conclusion}
\label{sec:conclusion}

We have extended Presto to benefit from GPU acceleration, validating three key hypotheses: (1) data must be read directly into GPU memory, (2) it must remain there throughout query execution, and (3) distributed queries require GPU-native inter-node communication. These principles underlie our careful integration of standard libraries---cuDF for GPU operators, UCX for GPU-to-GPU communication, and KvikIO for storage-to-GPU transfers---into Presto via Velox. Our implementation enables distributed Presto queries to execute entirely on GPU clusters without CPU/GPU memory transfers.

Our evaluation demonstrates up to 6$\times$ better cost/performance compared to CPU Presto for analytical workloads. We have demonstrated both scale-up (to 30TB datasets on 40 GPUs) and scale-out (adding distributed workers across machines), and the system is starting to run customer production workloads. Because our changes are integrated into open-source Velox, other Velox-based engines such as Spark/Gluten inherit the same GPU support.


\bibliographystyle{ACM-Reference-Format}
\bibliography{sample}



\end{document}